\documentclass[sigconf, anonymous=false, authorversion=true, nonacm=true]{acmart}

\pdfoutput=1

\usepackage{adjustbox}
\usepackage{multirow}
\usepackage{makecell}
\usepackage{enumitem}
\usepackage{xspace}

\AtBeginDocument{%
  \providecommand\BibTeX{{%
    \normalfont B\kern-0.5em{\scshape i\kern-0.25em b}\kern-0.8em\TeX}}}

\setcopyright{acmcopyright}
\copyrightyear{2020}
\acmYear{2020}
\acmDOI{}

\acmConference[SIGIR '20]{43rd International ACM SIGIR Conference on Research and Development in Information Retrieval}{July 25--30, 2020}{Xi'an, China}
\acmBooktitle{SIGIR '20: 43rd International ACM SIGIR Conference on Research and Development in Information Retrieval, July 25--30, 2020, Xi'an, China}
\acmPrice{15.00}
\acmISBN{}

\newcommand{\cast}{\textsc{CAsT}\xspace}
\newcommand{\msmarco}{\textsc{MS-MARCO}\xspace}
\newcommand{\ft}{$FT$\xspace}

\newcommand{\pretty}[1]{\textsf{#1}}
\newcommand{\query}{\pretty{Query}\xspace}
\newcommand{\plainu}{\pretty{Plain Utterance}\xspace}
\newcommand{\firstq}{\pretty{First Query}\xspace}
\newcommand{\ctxq}{\pretty{Context Query}\xspace}
\newcommand{\firstt}{\pretty{First Topic}\xspace}
\newcommand{\topics}{\pretty{Topic Shift}\xspace}
\newcommand{\ctx}{\pretty{Context}\xspace}
\newcommand{\coref}{\pretty{CoRef}\xspace}
\newcommand{\corefone}{\pretty{CoRef}$_1$\xspace}
\newcommand{\coreftwo}{\pretty{CoRef}$_2$\xspace}
\newcommand{\manual}{\pretty{Manual Utterance}\xspace}

\newcommand{\baseline}{\pretty{\cast baseline}\xspace}

\newcommand{\ts}{$TS$\xspace}

\newcommand{\tpe}{$TP_e$\xspace}
\newcommand{\tpi}{$TP_i$\xspace}
\newcommand{\cp}{$CP$\xspace}

\begin{document}


\title{Topic Propagation in Conversational Search}
\author{Ida Mele}
\affiliation{%
\institution{ISTI-CNR}}
\email{ida.mele@isti.cnr.it}

\author{Cristina Ioana Muntean}
\affiliation{%
\institution{ISTI-CNR}}
\email{cristina.muntean@isti.cnr.it}

\author{Franco Maria Nardini}
\affiliation{%
\institution{ISTI-CNR}}
\email{francomaria.nardini@isti.cnr.it}

\author{Raffaele Perego}
\affiliation{%
\institution{ISTI-CNR}}
\email{raffaele.perego@isti.cnr.it}

\author{Nicola Tonellotto}
\affiliation{%
  \institution{University of Pisa}}
  \email{nicola.tonellotto@unipi.it}

\author{Ophir Frieder}
\affiliation{%
  \institution{Georgetown University}}
\email{ophir@ir.cs.georgetown.edu}

\renewcommand{\shortauthors}{Mele et al.}



\begin{abstract}

In a conversational context, a user expresses her multi-faceted information need as a sequence of natural-language questions, i.e., utterances. Starting from a given topic, the conversation evolves through user utterances and system replies. The retrieval of documents relevant to a given utterance in a conversation is challenging due to ambiguity  of natural language and to the difficulty of detecting possible topic shifts and semantic relationships among utterances.
We adopt the 2019 TREC Conversational Assistant Track (\cast)
framework to experiment with a modular architecture performing: (i) topic-aware utterance rewriting, (ii) retrieval of candidate passages for the rewritten utterances, and (iii) neural-based re-ranking of candidate passages. We present a comprehensive experimental evaluation of the architecture assessed in terms of traditional IR metrics at small cutoffs. Experimental results show the effectiveness of our techniques that achieve an improvement up to $0.28$ (+93\%) for P@1 and $0.19$ (+89.9\%) for nDCG@3 w.r.t. the \cast baseline.
\end{abstract}





\maketitle


\section{Introduction}\label{sec:intro}

The increasing popularity of personal assistant systems has drawn attention to conversational information retrieval (IR) systems. 
These systems help users in several  activities (e.g., checking the weather forecast, performing e-commerce transactions).
This flexibility comes from the easiness with which a user can fulfill her information needs by simply conversing with the system as in a multi-turn conversation which evolves through utterances and replies.

Answering user utterances in a multi-turn conversation is not straightforward since the system must understand the questions, find relevant documents, and sort them based on their relevance to return a narrowed list of results (sometimes only one).
The utterances are formulated in natural language, so they are prone to ambiguity, polysemy of words, presence of acronyms, mistakes, and grammar misuse. More importantly, often a complex information need cannot be resolved with a single question, rather the user formulates multiple subsequent utterances related to each other.
Consider the following search conversation: (1) ``Tell me about the Neverending Story film.'', (2) ``What is it about?'', and (3) ``What are the main themes?''.
Only the first utterance is self-explanatory and easy to process by IR systems, while the others contain pronouns that refer to the previous topic (i.e., 2) or even they do not have any explicit reference to the conversation subject (i.e., 3).

We focus on rewriting user utterances to semantically enrich those utterances that lack context. We experiment with different automatic utterance rewriting techniques, showing that the enrichment of utterances with keywords enclosing the conversational topic yields to utterances similar to the manually re-formulated ones. These automatically rewritten utterances can be effectively processed by IR systems, allowing high precision results.
The problem of utterance rewriting was addressed by Ren et al.~\cite{Ren:WWW2018}. They proposed a sequence-to-sequence model for context-aware rewriting of conversational queries.
Other authors tried to identify the utterances relevant to a given turn~\cite{aliannejadi2019harnessing} or to predict the next question in the conversation~\cite{yang2017neural}. A different line of research focuses on improving the result ranking by incorporating external knowledge~\cite{Yang:SIGIR2018}. Conversational IR challenges are somehow related to those of Web search where the queries may be ambiguous or not-well formulated. Thus, many identify the user intent behind a query, mostly exploiting search logs for improving query suggestion~\cite{Cao:KDD2008}. Rather than leveraging search history, we focus on one information-seeking conversation at-a-time.

We improve the effectiveness at small cutoffs of IR systems in the context of multi-turn conversational searches through a modular architecture performing: (i) automatic utterance rewriting, (ii) first-stage retrieval of candidate passages for the rewritten utterances, and (iii) neural-based re-ranking of candidate passages.
The first step (\textit{utterance rewriting}) consists of enriching the current utterance with context, adding keywords that explicitly refer to the topic of the conversation or to the specific facet of the search.
The second step (\textit{first-stage retrieval}) narrows the search space by retrieving a limited set of passages relevant to the expanded utterance. The third step (\textit{neural-based re-ranking}) exploits a contextualized language model based on BERT to re-rank the retrieved passages~\cite{nogueira2019passage}.

Using the 2019 TREC Conversational Assistant Track (\cast) framework, we present a comprehensive experimental evaluation of our modular architecture in terms of IR metrics (e.g, P@1, P@3, and nDCG@3). Our results show that if an explicit reference to the conversational topic is missing, adding topic-related information is crucial for a better retrieval of the relevant passages. We also note that the conversation topic is typically enclosed in the first utterance. However, topic switches are also common in multi-turn conversations, so it is important to identify them for improving the quality of the results. Our techniques for utterance rewriting take into account all these variations, overcoming other approaches based on co-referencing~\cite{Gardner2017AllenNLP}. Experimental results show that our best approach achieves an improvement  of $0.25$ (+113\%) for P@1 at the first-stage retrieval and of $0.28$ (+93\%) at the neural-based re-ranking stage  w.r.t. the \cast provided baseline. The improvement for nDCG@3 is instead of  $0.19$ (+89.9\%).


\section{Utterance Rewriting}\label{sec:utt_rewriting}

We use the term \textit{topic} to indicate the theme of a multi-turn conversation and \textit{utterance} for the natural-language request formulated by the user in one turn. An utterance is later converted into a \textit{query} which is given as input to a retrieval system. 

To retrieve highly relevant passages, our rewriting approach  aims to maintain the conversation context since natural-language utterances might be not self-explanatory. Indeed, if these utterances are used directly as queries, they can be ambiguous or too generic.

The conversational search utterances express different facets of the user information need. An utterance can be self-explanatory or vague, it can explore several aspects regarding an implicit topic, can be a specialization, a paraphrase, and even a topic switch. All these nuances are hard to capture and even harder to resolve~\cite{Ren:WWW2018}. To improve the expressiveness of corresponding queries, we propose utterance rewriting techniques.
Our techniques transform potentially vague and ambiguous conversational utterances to effectively-processed, self-explanatory queries for automated retrieval.

Part-of-speech tagging, named entity resolution, dependency parsing and co-reference resolution are the linguistic analysis components which help identify significant pieces of text and which we exploit in our utterance rewriting strategies. We extract various linguistic features from the utterances\footnote{By using the \textsc{spacy} library available at \url{https://spacy.io/usage/linguistic-features}.} and develop core modules, each module exploiting a distinct set of linguistic features:

\begin{itemize}[leftmargin=*,topsep=2pt]
    \item \textit{First Topic} (\ft) uses dependency parsing to identify significant noun chunks, e.g., objects or subjects differing from pronouns. We apply this module to the first utterance of each conversation to identify its topic.
    \item \textit{Co-reference Resolution} (\textit{CR}) finds all expressions that refer to the same entity in a text. We experiment with two versions of co-referencing: (1) the AllenNLP co-referencing model~\cite{Gardner2017AllenNLP, Lee2017EndtoendNC}, and (2) the \textsf{neuralcoref} model from the Transformers library\footnote{\url{https://github.com/huggingface/neuralcoref}}.
    \item \textit{Topic Shift} (\ts) is based on cues (e.g., ``tell me about'') and tries to capture a topic change that might be significant for the conversation from that point on. It is based on dependency parsing for extracting the \textit{current topic} from the utterance.    \item \textit{Context Propagation} (\cp) captures all possible noun chunks, different from pronouns, which are either subjects or objects, to enrich the context of the conversation up to the current turn.
\end{itemize}

We explore different combinations of these modules to determine the best methodology for the enrichment of conversational utterances. Due to space limitations, we present here the combinations of modules for which we discuss our findings in Sec.~\ref{sec:experiments}. 
\begin{itemize}[leftmargin=*,topsep=2pt]
    \item \firstt, a combination of \ft, \tpe, and \tpi. We propagate the first topic throughout the entire conversation addressing the cases of replacing the explicit pronouns as well as the implicit references to the conversation topic.
    \item \topics, a combination of \ft, \tpe, \tpi, and \ts. This combination is similar to the previous one (\firstt) but propagates the current topic, instead of the first one, if a topic shift is detected.
    \item \ctx, a combination of \ft, \tpe, \ts, and \cp. We use noun chunks which are subjects or objects from all previous utterances to keep track of the entire context up to the current turn. In this way, the utterance is enriched with both topics and context.     \item \coref. We rewrite utterances using \textit{CR}: the AllenNLP model (\corefone) or the Transformers Neural Coref  (\coreftwo).
\end{itemize}


\section{Experimental Setup}\label{sec:exp_setup}

We use the 2019 TREC Conversational Assistant Track (CAsT) dataset \footnote{http://www.treccast.ai/} that includes three collections: (1) TREC CAR (Complex Answer Retrieval) containing $\sim29M$ of passages extracted from approximately $5M$ Wikipedia articles, (2) \msmarco (MAchine Reading COmprensation) made of $\sim8M$ passages from answer candidates of Bing search engine, and (3) WAPO (WAshington POst) dataset consisting of $\sim8M$ passages extracted from $\sim608K$ news articles. 
The \cast dataset also provides $80$ search conversations ($30$ for training and $50$ for evaluation), each having from $8$ to $12$ utterances. 
For $194$ utterances from evaluation conversations, \cast provides also utterance-passage relevance judgements graded on a three-point scale, i.e., $2$ very relevant, $1$ relevant, and $0$ not relevant.

\noindent \textbf{Metrics}. We evaluate the effectiveness of our system with traditional IR metrics:
Mean Average Precision (MAP), Mean Reciprocal Rank (MRR), normalized Discounted Cumulative Gain and Precision for cutoffs at 1 and 3 (nDCG@3, P@1, and P@3). The use of such small cutoffs is common for the conversational IR task since the user 
expects to receive one crisp answer on the top of the list rather than a long list of potentially relevant results.

\noindent \textbf{First-stage retrieval}.
For indexing and querying the \cast dataset, we use Indri\footnote{https://www.lemurproject.org/indri.php}. We index  the three datasets by removing stopwords. We experiment with and without stemming, and we observe better results with the \textit{Krovetz} stemmer. We also experiment with different Indri querying methods (e.g., TF-IDF, BM25, inQuery). We achieve the best retrieval performance with the Indri language model based on Dirichlet smoothing with parameter $\mu = 2500$. We also apply pseudo-relevance feedback (PRF) based on the RM3 algorithm~\cite{Clark:2010} using $20$ keywords taken from the top-$20$ results and mixing parameter $\gamma = 0.5$.

\noindent \textbf{Neural re-ranking}.
Regarding the third stage, we use the model proposed by Nogueira and Cho \cite{nogueira2019passage}. The model fine-tunes the BERT base pre-trained model for re-ranking on \msmarco passage retrieval dataset.
For each query Indri retrieves $1000$ results but we select the top $200$ documents only to feed the  re-ranking step. Consistently with ~\cite{DBLP:journals/corr/abs-1904-12683} smaller and larger cutoffs ($50$ and $1000$ documents) provide worse end-to-end performance.

\noindent \textbf{Competitors}.
\cast provides a baseline consisting  in queries generated from  utterances by applying stopword removal and AllenNLP co-reference resolution. An  Indri run of these queries with query likelihood and no PRF is our  \baseline. To provide a consistent experimental setting,  we instead process the \cast queries (hereinafter \query),  our rewriting modules (see Section \ref{sec:utt_rewriting}) and the following competitors by   using PRF and the  Indri language model: 
 \begin{itemize}[leftmargin=*,topsep=2pt]
    \item \firstq: given a conversation, the current query, $q_i$, is expanded with the first-turn query, $q_1$. This is done to perform a simple query rewriting (e.g., $q_1$~+~$q_i$).
    \item \ctxq: given a conversation, the current query is enriched with the first query and the one appearing in the previous turn (e.g., $q_1$~+~$q_{i-1}$~+~$q_i$).
    \item \plainu: utterances provided by \cast which represent the original user requests, i.e., without performing any rewriting.
    \item \manual: utterances manually rewritten by human annotators (provided by \cast). The result of this run can be considered an upper bound as manual rewriting should ideally outperform any automatic system.
\end{itemize}

\section{Experimental Results}
\label{sec:experiments}

We perform an experimental evaluation to answer the following research questions (RQs):
\begin{itemize}[leftmargin=*,topsep=2pt]
    \item RQ1:  To what extent is our rewritten utterances beneficial for first-stage retrieval and re-ranking of passages? Are some rewriting techniques better than others?
    \item RQ2: What is the impact of the rewritten utterances only on re-ranking?
    \item RQ3: Shall we use different rewriting techniques for the two different stages: first-stage retrieval and re-ranking?
  \end{itemize}

To answer to RQ1, we evaluated the performance of the first-stage retrieval only and of the whole end-to-end system. The results are shown in Table~\ref{tb:all_results}.
\begin{table*}[t!]
\begin{small}
 \centering
 \caption{Retrieval and re-ranking results using different inputs. Best results are shown in bold.  We highlight statistical significant differences w.r.t. the winning method \topics with $\blacktriangledown$ and $\blacktriangle$ for $p<0.01$ according to the two-sample t-test. \manual is not a method, although shows the best results, since utterances were manually re-written by human annotators.\vspace{-2mm}}
 \label{tb:all_results}
 \begin{adjustbox}{max width=\textwidth}
 \begin{tabular}{lcccccccccc}
 \toprule
  & \multicolumn{5}{@{}c@{}}{\makecell{{First-stage Retrieval}}} & \multicolumn{5}{@{}c@{}}{\makecell{{Neural Re-ranking}}}\\
 \cmidrule(lr){2-6}\cmidrule(lr){7-11}
  & {MAP} & {MRR} & {nDCG@3} & {P@1} & {P@3} & {MAP} & {MRR} & {nDCG@3}  & {P@1} & {P@3}\\
 \midrule

 \baseline & 0.1299$\blacktriangledown$ & 0.3178$\blacktriangledown$ &  0.1477$\blacktriangledown$ &  0.2254$\blacktriangledown$ &  0.2428$\blacktriangledown$    & 0.1316$\blacktriangledown$  & 0.4002$\blacktriangledown$  & 0.2089$\blacktriangledown$  & 0.3064$\blacktriangledown$ & 0.2987$\blacktriangledown$ \\
 \midrule
 \query  & 0.1741$\blacktriangledown$ & 0.4216$\blacktriangledown$ &  0.2181$\blacktriangledown$ &  0.3353$\blacktriangledown$ &  0.3218$\blacktriangledown$  & 0.1666$\blacktriangledown$ &  0.4723$\blacktriangledown$ &  0.2660$\blacktriangledown$ & 0.3547$\blacktriangledown$ &   0.3624$\blacktriangledown$ \\

 \firstq & 0.2091  & 0.4931  & 0.2663  & 0.3988  & 0.4008 & 0.2095 & 0.5945$\blacktriangledown$  & 0.3195$\blacktriangledown$  & 0.4884$\blacktriangledown$  & 0.4477$\blacktriangledown$\\
  \ctxq & 0.1903$\blacktriangledown$ & 0.4709$\blacktriangledown$ & 0.2315$\blacktriangledown$ & 0.3565 & 0.4008$\blacktriangledown$  & 0.1905$\blacktriangledown$ & 0.5624$\blacktriangledown$ & 0.2997$\blacktriangledown$ & 0.4535$\blacktriangledown$ & 0.4147$\blacktriangledown$ \\
  \midrule
  \plainu & 0.1416$\blacktriangledown$  & 0.3483$\blacktriangledown$  & 0.1735$\blacktriangledown$  & 0.2849$\blacktriangledown$  & 0.2694$\blacktriangledown$ &  0.1408$\blacktriangledown$  & 0.4605$\blacktriangledown$  & 0.2791$\blacktriangledown$  & 0.3953$\blacktriangledown$  & 0.3682$\blacktriangledown$ \\
  {\corefone} & 0.1691$\blacktriangledown$ & 0.4203$\blacktriangledown$ & 0.2148$\blacktriangledown$ &  0.3410 & 0.3218$\blacktriangledown$ &  0.1797$\blacktriangledown$ & 0.5563$\blacktriangledown$ & 0.3474 & 0.4795 & 0.4405$\blacktriangledown$ \\
  {\coreftwo} & 0.1845$\blacktriangledown$ & 0.4308$\blacktriangledown$ & 0.2209$\blacktriangledown$ & 0.3526 & 0.3430$\blacktriangledown$ &  0.1906$\blacktriangledown$ & 0.5718$\blacktriangledown$ & 0.3483 & 0.5000 & 0.4554$\blacktriangledown$ \\
  \midrule
  \ctx & 0.2053 & 0.4701$\blacktriangledown$ & 0.2401$\blacktriangledown$ & 0.3757 & 0.3680$\blacktriangledown$ &  0.2098  & 0.6315 & 0.3692 & 0.5291 & 0.4961\\ 
  \firstt & \textbf{0.2286} &  0.5543 & \textbf{0.3041} &  0.4682 & \textbf{0.4644} &  \textbf{0.2335} & 0.6763 & 0.3897 & \textbf{0.5954} & 0.5453\\
  \topics & 0.2274 & \textbf{0.5558} & \textbf{0.3041} & \textbf{0.4798} & 0.4605 & 0.2330  & \textbf{0.6801} & \textbf{0.3967} & 0.5872 & \textbf{0.5523}\\
  \midrule
  \manual & 0.2978$\blacktriangle$ & 0.6859$\blacktriangle$ & 0.3828$\blacktriangle$ & 0.6127$\blacktriangle$ & 0.5491$\blacktriangle$ &  0.3055$\blacktriangle$ &  0.8006$\blacktriangle$ & 0.5167$\blacktriangle$ & 0.7168$\blacktriangle$ & 0.6763$\blacktriangle$\\
 \bottomrule
 \end{tabular}
 \end{adjustbox}
 \end{small}
 \end{table*}

The worst performance is achieved by the \baseline as oftentimes the co-referencing fails to enrich the query and after stopword removal some of the queries become even empty. We  note that, after the re-ranking stage, there is a performance improvement ($0.08$ of P@1).
\plainu performs only slightly better than the \baseline, which is expected, as many utterances are not self-expressive and do not contain the topic of the conversation (e.g.,~``What is it about?''). 

We improve the \baseline by using PRF at the first-stage retrieval (\query); the overall performance of \query improves (e.g.,~P@1 improves by $0.11$ at the first-stage retrieval and of $0.05$ after re-ranking). This result confirms that  adding keywords to the original query by using PRF is beneficial.

We also tried to enrich the queries with more targeted keywords taken from the conversations. In particular, we observed that an improvement can be achieved by simply adding the keywords of the first query without (\firstq) or with (\ctxq) the context of the previous one. In particular, in terms of P@1, the \firstq method improves the \baseline by $0.17$ ($+77\%$) at the first-stage retrieval and by $0.18$ ($+60\%$) after re-ranking.

The use of co-referencing on plain utterances (i.e., \corefone and \coreftwo) is insufficient as the co-referencing approaches may miss the conversational topic.
We also noticed that the results for \baseline are weaker than those for \corefone although the processes for generating the queries are similar (they both use AllenNLP). However, for \corefone we used PRF in the first-stage retrieval, while the \baseline is based only on query likelihood.

Regarding our utterance rewriting approaches, the best performance is achieved by \firstt and \topics. The last approach is best as it not only captures the main topic of the conversation (e.g., from the first utterance) but also adjusts it when needed (e.g., if there is a change of the topic). As an example, the user might start searching for ``lung cancer'' and then switch to ``throat cancer''. The improvement of \topics over the baseline is $0.25$ ($+113\%$) for P@1 for first-stage retrieval and $0.28$ ($+93\%$) for the end-to-end system. Indeed, \firstt shows similar results. On the other hand, enriching utterances with the context of all the previous utterances (\ctx) worsens the performance as adding too many search keywords helps recall more than precision.

Another interesting finding relates to the re-ranking impact on the final result list. The re-ranking stage improves the performance for every method compared to first-stage retrieval, for all metrics except MAP, making it a fundamental part of the system.  Our \firstt and \topics methods yield the best end-to-end system performance, confirming the importance of a well-written utterance, e.g., with the \topics rewriting technique, P@1 goes from $0.48$ to $0.59$ with an improvement of $0.11$ ($+22\%$).

From these experiments, we conclude that both PRF and neural-based re-ranking boost the performance of a conversational IR system. However, they are more beneficial with automatic rewritten utterances. Indeed, given the \cast queries, we improve the \baseline results leveraging PRF and neural-based re-ranking, but the improvement is small (as shown by the comparison between the \baseline and the \query results). However, performance is consistently enhanced thanks to re-written utterances used as input at both stages as corroborated by the comparison between \baseline and \topics.
Additionally, our \topics approach achieves performance close to the one registered with \manual. For example, for P@1 the distance between \topics and \manual is $0.14$ ($0.13$ after re-ranking) while between the \baseline and \manual is much larger, $0.39$ ($0.41$ after re-ranking), so the overall improvement is up to $0.28$ for the end-to-end system.

To answer RQ2, we replaced the BERT input queries with different rewritten utterances and we used as input passages the \baseline results of the first stage.
As an example, the input  query-passage pair for BERT, obtained from the \baseline, would have as textual equivalent: \texttt{<"tiger sharks", "Sand tiger sharks are often the targets of scuba divers who wish to observe or photograph these animals">}. We can replace \texttt{"tiger sharks"} with \texttt{"Tell me more about tiger sharks. the different types sharks"} and use the new pair as input for BERT. 
This illustrates that even with a low-quality result list obtained at the first-stage retrieval, we can use improved utterances at the re-ranking stage to achieve good rankings.
In Table~\ref{tb:baseline}, we report the end-to-end performance of a system using the \baseline at the first-stage retrieval and the BERT re-ranking using different kinds of input queries.
By using the \baseline as input for both stages, P@1 with re-ranking has only a marginal  improvement of $0.08$, as reported also in Table~\ref{tb:all_results}. Better results are obtained when using the \plainu (without any rewriting). This confirms that BERT re-ranking, having been trained on questions and passage pairs, works better when utterances rather than query keywords are used as input~\cite{nogueira2019passage}.
The best performing method is once more \topics with a much larger improvement w.r.t. the baseline, e.g., for P@1 is $0.31$ (from $0.22$ to $0.53$). Moreover, \topics performance are very close to the ones of \manual for all  metrics.
The experiments on RQ2 highlight the importance of having well-formed utterances even for neural re-ranking. The richer context introduced in the rewritten utterances results in fact to largely help  BERT to improve the ranking quality.

\begin{table}[htb]
\caption{BERT re-ranking over the \baseline.\vspace{-2mm}}\label{tb:baseline}
 \begin{adjustbox}{width=\columnwidth,totalheight={\textheight},keepaspectratio}
 \begin{tabular}{lccccc}
 \toprule
& {MAP} & {MRR} & {nDCG@3}  & {P@1} & {P@3}  \\
 \midrule
\baseline  & 0.1316  & 0.4002  & 0.2089  & 0.3064 & 0.2987 \\
\plainu & 0.1539  & 0.4193 & 0.2538 & 0.3179 &  0.3526\\
\topics & \textbf{0.1964}  & \textbf{0.6076} & \textbf{0.3635} &  \textbf{0.5318 }&  \textbf{0.4913}\\
\midrule
\manual & 0.2038 & 0.6192 & 0.3853 & 0.5318 & 0.5202\\
  \bottomrule
\end{tabular}
\end{adjustbox}
\vspace{-2mm}
\end{table}

To answer RQ3, we select for the first-stage retrieval the method with the best recall@200. We use its result lists together with utterances obtained by different rewriting techniques as input for the re-ranking stage.  Using this approach, we verify if a higher number of candidate positive results is effectively exploited in the re-ranking stage. We observed that the best method in terms of recall is \ctx with a recall@200 of $0.4967$. This is expected since the context keywords add much information to the query and consequently increase the recall of the result list. Due to space limitations, we show only the experiment where \ctx results are used together with utterances rewritten with the best performing method for re-ranking (\topics). 
The end-to-end performance with  \ctx-\topics is respectable: MAP = 0.2178, MRR = 0.6645, nDCG@3 = 0.3925, P@1 = 0.5872, and P@3= 0.5291. Anyway, they do not top \topics used at both stages since using context at the first-stage retrieval improves recall but may worsen the precision of the top 200 results. The experiments on RQ3 show that different rewriting techniques can be used at different steps, and each stage of the IR system can be tuned or improved separately.


\section{Conclusion}
\label{sec:concl}
We proposed a three-steps architecture for conversational search and, to this regard, we developed several utterance rewriting techniques.
We experimentally showed that well-formed and self-ex\-pres\-sive utterances can improve the precision of results consistently, effect which is propagated in both retrieval stages. The performance of our best rewriting technique, based on detecting the conversational topic as well as its possible variations (\topics), is the closest to the manually rewritten utterances w.r.t. other methods. 

\vspace{-0.1cm}
\begin{acks}
Work partially supported by the Italian Ministry of Education and Research (MIUR) in the framework of the CrossLab project (Departments of Excellence).  Work partially supported by the BIGDATAGRAPES project funded by the EU Horizon 2020 research and innovation programme under grant agreement No. 780751, and by the OK-INSAID project funded by the Italian Ministry of Education and Research (MIUR) under grant agreement No. ARS01\_00917.
\end{acks}

\bibliographystyle{ACM-Reference-Format}
\bibliography{biblio}

\end{document}